\documentclass[cup6b]{cupbook}
\usepackage{graphicx}
\usepackage{natbib}
\title[Rome, Italy, 27--30 April 2009]
      {The coming of age of X-ray polarimetry}
\author{}
\date{}
\begin{document}
\pagenumbering{arabic}

\newcommand{\go}{\mathrel{\raise.3ex\hbox{$>$}\mkern-14mu
             \lower0.6ex\hbox{$\sim$}}}
\newcommand{\lo}{\mathrel{\raise.3ex\hbox{$<$}\mkern-14mu
             \lower0.6ex\hbox{$\sim$}}}
\newcommand{\veck}{{\bf k}}
\newcommand{\vecB}{{\bf B}}
\newcommand{\vecmu}{{\mbox{\boldmath $\mu$}}}

\author[D. Lai et al.]
{Dong Lai (Cornell University) 
\and Wynn C.G. Ho (University of Southampton) 
\and Matthew van Adelsberg (Kavli Institute for Theoretical Physics, UCSB)
\and Chen Wang (NAOC and Cornell University)
\and Jeremy S. Heyl (University of British Columbia)}
\chapter{Polarized X-rays from Magnetized Neutron Stars}

\abstract{
We review the polarization properties of X-ray emission from
highly magnetized neutron stars, focusing on emission from the stellar
surfaces. We discuss how x-ray polarization
can be used to constrain neutron star magnetic field and emission geometry,
and to probe strong-field quantum electrodynamics
and possibly constrain the properties of axions.}

\section{Introduction}

One of the most important advances in neutron star (NS) astrophysics in the
last decade has been the detection and detailed studies of surface (or
near-surface) X-ray emission from a variety of isolated NSs 
\cite{kaspi06,harding06}.
This has been made possible by X-ray telescopes such as 
{\it Chandra} and {\it XMM-Newton}. Such studies can potentially provide
invaluable information on the physical properties and evolution of NSs
(e.g., equation of state at super-nuclear densities, cooling history,
surface magnetic fields and compositions, different NS populations).
The inventory of isolated NSs with detected surface emission includes:
(i) Radio pulsars: e.g., the phase-resolved spectroscopic observations 
of the ``three musketeers'' revealed the geometry of the NS polar caps;
(ii) Magnetars (AXPs and SGRs): e.g., the quiscent emission 
of magnetars consists of a blackbody at
$T\sim 0.5$~keV with a power-law component (index 2.7-3.5), plus
significant emission up to $\sim 100$~keV;
(iii) Central Compact
Objects (CCOs) in SNRs: these now include 6-8 sources, several have
$P,{\dot P}$ measurements and two have absorption lines \cite{deluca08};
(iv) Thermally-Emitting Isolated NSs:
these are a group of 7 nearby ($\lo 1$~kpc) NSs with low
($\sim 10^{32}$~erg~s$^{-1}$) X-ray luminosities and long (3-10~s)
spin periods, and recent observations have revealed 
absorption features in many of the sources \cite{vankerkwijk07,kaplan08}

In the coming decade, the most important goals in NS astrophysics 
include: (i) understanding how these different
types of NSs evolve and relate to each others;
(ii) elucidating the different observational manifestations (e.g.,
radiative processes in NS atmospheres and magnetospheres);
(iii) using NSs to probe physics under extreme conditions.
An obvious message of this paper is that in addition to imaging, timing and
spectroscopy, X-ray polarimetry provides a window to study NSs:
e.g., even when the spectrum or light curve is 
``boring'', polarization can still be interesting and very informative.
Recent advances in detector technology suggest that polarimetry study of
X-ray sources holds great promise in the future (see contributions
by E. Costa, J. Swank, and M. Weiskopf in this proceedings).

\section{Polarized X-rays from NSs: Basics}


The surface emission from magnetized NSs (with $B\go 10^{12}$\,G) is
highly polarized \cite{gnedin74,pavlov00}
for the following reason.  In the magnetized
plasma that characterizes NS atmospheres, X-ray photons propagate in
two normal modes: the ordinary mode (O-mode, or $\parallel$-mode) is
mostly polarized parallel to the $\veck$-$\vecB$ plane, while the
extraordinary mode (X-mode, or $\perp$-mode) is mostly polarized
perpendicular to the $\veck$-$\vecB$ plane, where $\veck$ is the
photon wave vector and $\vecB$ is the external magnetic field.
This description of normal modes applies under
typical conditions, when the photon energy $E$ is much less than the
electron cyclotron energy $E_{Be}=\hbar eB/(m_ec)=11.6\,B_{12}$\,keV
[where $B_{12}=B/(10^{12}\,{\rm G})$], $E$ is not too close to the ion
cyclotron energy $E_{Bi}=6.3\,B_{12}(Z/A)$\,eV (where $Z$ and $A$ are
the charge number and mass number of the ion), the plasma density is
not too close to the vacuum resonance (see below) and $\theta_B$ (the
angle between $\veck$ and $\vecB$) is not close to zero.  Under these
conditions, the X-mode opacity (due to scattering and absorption) is
greatly suppressed compared to the O-mode opacity, $\kappa_X\sim                
(E/E_{Be})^2\kappa_O$. 
As a result, the X-mode photons escape
from deeper, hotter layers of the NS atmosphere than the O-mode
photons, and the emergent radiation is linearly polarized to a high
degree \cite{pavlov00,ho01,ho03,vanadelsberg06}.
Thus, if we were to put a polarimeter on the NS surface
we would measure high-polarization X-rays.

To translate this ``surface measurement'' into the observed signals at
infinity, we need to understand how a photon (including its
polarization state) evolves as it travels from the emission point to
the observer. This will involve considering the geometry of the emission region
(including magnetic field), light bending, and polarization evolution.
Before discussing these issues, we summarize some of the 
general expected X-ray polarization characteristics:

(i) The X-ray polarization vector is either $\perp$ or $\parallel$ to the
$\veck$-$\vecmu$ plane, depending on the photon energy and 
surface field strength, even when the surface field is non-dipole!
(Here $\vecmu$ is the magnetic dipole axis.)

(ii) As the NS rotates, we will obtain a ``linear polarization sweep''
(as in the rotating vector model of radio pulsars). This will provide
a constraint on the dipole magnetic field geometry.

Thus, measurements of X-ray polarization, particularly when
phase-resolved and measured in different energy bands, could provide
unique constraints on the NS magnetic field (both the dipole component
and the ``total'' strength) and geometry. There is only a modest
dependence on $M/R$ of the NS, but as we discuss in the next section,
quantum electrodynamics (QED) plays an important role.

{\it Other Issues}: (i) For sufficiently low $T$ and high $B$, there is a
possibility that the NS surface is in a condensed (metallic) form, with
negligible ``vapor'' above it \cite{medin07}. This has been suggested 
in the case of the TINS RX J1856.5-3754 \cite{vankerkwijk07,ho07}.
Radiation from a condensed surface is certainly quite different from 
an atmosphere, with distinct X-ray polarization characteristics 
\cite{vanadelsberg05}. (ii) In the case of magnetars, Compton scatterings
by mildly relativistic e$^\pm$ in the NS magnetosphere/corona are important 
in determining the X-ray spectra at $E\go 2$~keV 
\cite{fernandez07,nobili08}. How such scatterings
affect the polarization signals of surface emission has not been studied.

\section{QED Effects on X-ray Polarization Signals}

It has long been predicted from quantum electrodynamics (QED) that in
a strong magnetic field the vacuum becomes birefringent 
\cite{adler71,heyl97}.
While this vacuum polarization effect
makes the photon index of refraction deviate from unity only when
$B\go 300 B_{\rm Q}$, where $B_{\rm Q}=m_e^2c^3/(e\hbar)=4.414\times
10^{13}$\,G is the critical QED field strength, it can significantly
affect the spectra of polarization signals from magnetic NSs in more
subtle way, at much lower field strengths 
In particular, the combined
effects of vacuum polarization and magnetized plasma gives rise to a
``vacuum resonance'': 
a photon may convert
from the high-opacity mode to the low-opacity one and vice verse when
it crosses the vacuum resonance region in the inhomogeneous NS
atmosphere \cite{lai02}.
For $B\go 7\times 10^{13}$\,G (see below), this vacuum resonance
phenomenon tends to soften the hard spectral tail due to the
non-greyness of the atmospheric opacities and suppress the width of
absorption lines, while for $B\lo 7\times 10^{13}$\,G, the spectrum is
unaffected \cite{ho03,ho04,lai03a,lai03b,vanadelsberg06}.

The QED-induced vacuum birefringence influences the X-ray polarization
signals from magnetic NSs in two ways.  (i) {\it Photon mode
conversion in the NS atmosphere:} Since the mode conversion depends
on photon energy and magnetic field strength, this vacuum resonance
effect gives rise to a unique energy-dependent polarization signal in
X-rays: For ``normal'' field strengths ($B\lo 7\times 10^{13}$\,G),
the plane of linear polarization at the photon energy $E\lo 1$\,keV is
perpendicular to that at $E\go 4$\,keV, while for ``superstrong''
field strengths ($B\go 7\times 10^{13}$\,G), the polarization planes
at different energies coincide \cite{lai03b,vanadelsberg06}.
(ii) {\it Polarization mode decoupling in the magnetosphere:}
The birefringence of the magnetized QED vacuum decouples the photon
polarization modes, so that as a polarized photon leaves the NS
surface and propagates through the magnetosphere, its polarization
direction follows the direction of the magnetic field up to a large
radius (the so-called polarization limiting radius). The result is
that although the magnetic field orientations over the NS surface may
vary widely, the polarization directions of the photon originating
from different surface regions tend to align, giving rise to large
observed polarization signals \cite{heyl02,vanadelsberg06,wang09}.

\subsection{QED Effect in NS Atmospheres}

\begin{figure}
\centering
\vskip -1.5cm
\includegraphics[scale=0.34]{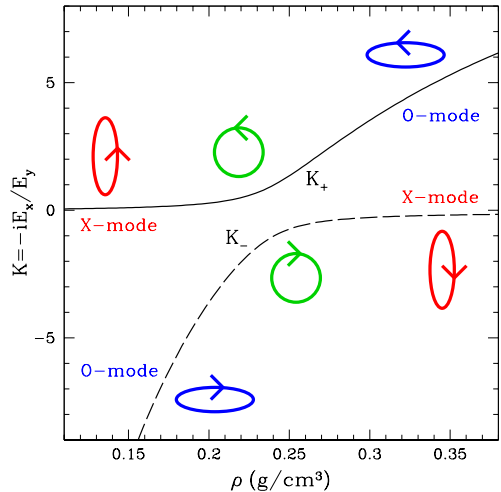}
\vskip -1.4cm
\caption{
Vacuum resonance in a NS atmospheres: The polarization ellipticities of
the two photon modes are shown as a function of density, for $B=10^{13}$~G,
$E=5$~keV, $Y_e=1$ and $\theta_B=45^\circ$.
}
\label{fig1}
\end{figure}

\begin{figure}
\centering
\vskip -0.5cm
\includegraphics[scale=0.44]{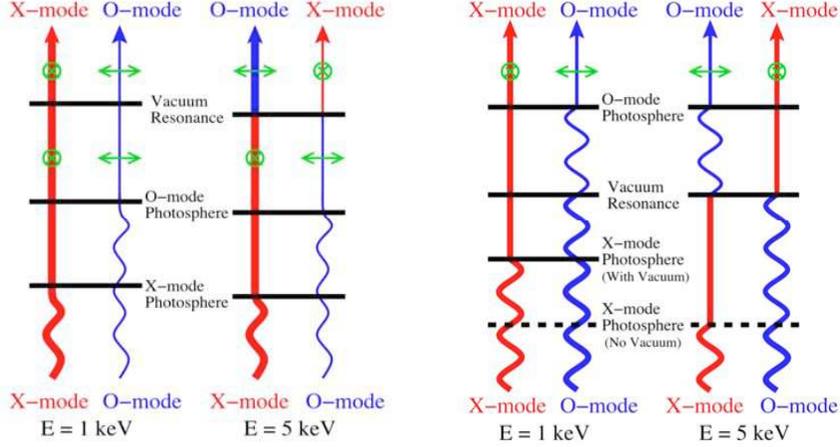}
\vskip -1.5cm
\caption{
A schematic diagram illustrating how vacuum polarization affects
the polarization state of the emergent radiation from a magnetized
NS atmosphere. The left panel is for 
$B\lo B_l\simeq 7\times 10^{13}$~G, and the right panel 
for $B\go B_l$. The photosphere is defined where the optical depth 
(measured from outside) is $2/3$.
}
\label{fig2}
\end{figure}

In a magnetized NS atmosphere, both the plasma and vacuum polarization
contribute to the dielectric tensor of the medium \cite{gnedin78,meszaros79}.
The vacuum polarization
contribution is of order $10^{-4}(B/B_Q)^2f(B)$ 
(where $f\sim 1$ is a slowly varying function of $B$), and is quite small 
unless $B\gg B_Q$. The plasma contribution depends on $(\omega_p/\omega)^2
\propto \rho/E^2$.
The ``vacuum''  resonance arises when the effects of
vacuum polarization and plasma on the polarization of the photon modes
``compensate'' each other. For a photon of energy $E$ (in keV), the vacuum
resonance occurs at the density
\begin{equation}
\rho_V\simeq 0.964\,Y_e^{-1}B_{14}^2E^2 f^{-2}~{\rm g~cm}^{-3},
\end{equation}
where $Y_e$ is the electron fraction \cite{lai02}.
Note that $\rho_V$ lies in the range of the typical densities
of a NS atmosphere.
For $\rho\go\rho_V$ (where the plasma effect dominates the dielectric
tensor) and $\rho\lo\rho_V$ (where vacuum polarization dominates), the
photon modes are almost linearly polarized --- they are the usual
O-mode and X-mode described above; at $\rho=\rho_V$, however, both
modes become circularly polarized
as a result of the ``cancellation'' of the plasma and vacuum polarization
effects (Fig.~\ref{fig1}). 
When a photon propagates outward
in the NS atmosphere, its polarization state will evolve adiabatically
if the plasma density variation is sufficiently gentle. Thus the photon
can convert from one mode into another as it traverses the
vacuum resonance. For this conversion to be effective,
the adiabatic condition must be satisfied:
\begin{equation}
E\go E_{\rm ad}=1.49\,\bigl(f\,\tan\theta_B |1-u_i|\bigr)^{2/3}
\left({5\,{\rm cm}/H_\rho}\right)^{1/3}~{\rm keV},
\label{condition}
\end{equation}
where $\theta_B$ is the angle between ${\veck}$ and ${\vecB}$, $u_i           
=(E_{Bi}/E)^2$ ($E_{Bi}$ is the ion cyclotron energy),
and $H_\rho=|ds/d\ln\rho|$ is the density scale
height (evaluated at $\rho=\rho_V$) along the ray.
For a typical atmosphere density scale height ($\sim 1$~cm), 
adiabatic mode conversion requires $E\go 1$-2~keV \cite{lai02,lai03a}.

The location of vacuum resonance relative to the photospheres of X-mode 
and O-mode photons are important. For magnetic field strengths satisfying
\cite{lai03a,ho04}
\begin{equation}
B \go B_l \simeq 6.6\times 10^{13}\, T_6^{-1/8}E_1^{-1/4}S^{-1/4}\mbox{ G},
\label{eq:maglim}
\end{equation}
where $T_6=T/(10^6~{\rm K})$ and $S=1-e^{-E/kT}$,
the vacuum resonance density lies between the X-mode and O-mode
photospheres for typical photon energies, leading to suppression of
spectral features and softening of the hard X-ray tail characteristic
of the atmospheres. For ``normal'' magnetic fields, $B\lo B_l$, the
vacuum resonance lies outside both photospheres, and the emission
spectrum is unaltered by the vacuum resonance, although the observed
polarization signals are still affected \cite{lai03b}.
See Fig.~\ref{fig2}.

\begin{figure}
\centering
\vskip -0.2cm
\includegraphics[scale=0.35]{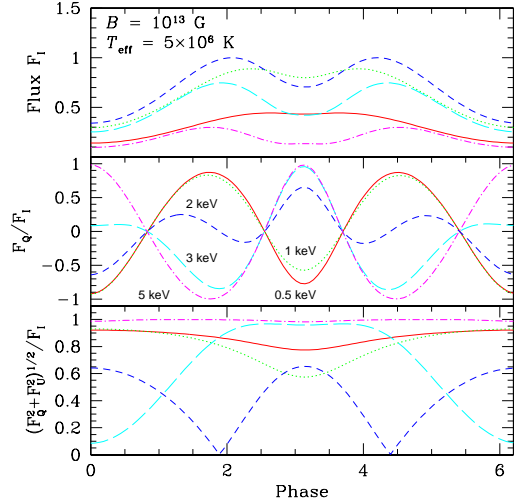}
\vskip -0.3cm
\caption{
Lightcurve and polarization as a function of rotation phase for a
NS hot spot with $B=10^{13}$ G, $T_{\rm eff}=5\times 10^6$ K.
The angle of the spin axis relative to the line of sight is
$\gamma=30^\circ$, and the inclination of the magnetic axis relative to
the spin axis is $\eta=70^\circ$.  
}
\label{fig3}
\end{figure}


\begin{figure}
\centering
\vskip -0.2cm
\includegraphics[scale=0.35]{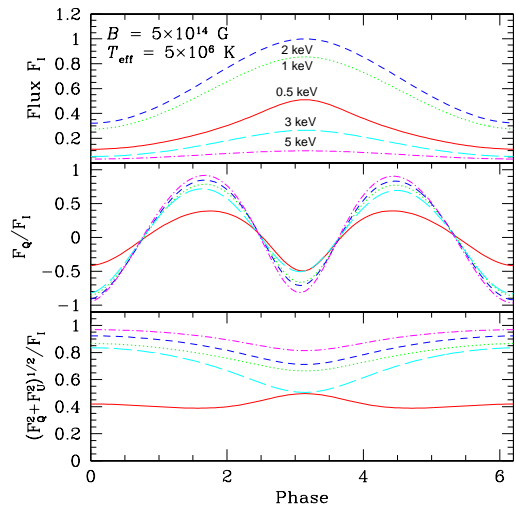}
\vskip -0.3cm
\caption{
Same as Fig.~\ref{fig3}, but for $B=5\times 10^{14}$~G.
}
\label{fig4}
\end{figure}

Figures \ref{fig3}-\ref{fig4} 
give some examples of the polarization signal of a NS 
hot spot. We see that for $B\lo B_l$, the sign of the $F_Q$
Stokes parameter is opposite for low and high energy photons; this
implies that the planes of polarization for low and high energy
photons are perpendicular. For $B\go B_l$, the planes of linear 
polarization at different $E$'s coincide.
This is a unique signature of vacuum polarization.

\subsection{QED Effect in Magnetospheres: Polarization Evolution}

Consider radiation from a large patch of the NS, with $\vecB$ varying
significantly across the emission region. Recall that locally at the NS
surface, the emergent radiation is dominated by one of the two modes.
If the photon polarization were parallel-transported to infinity, then the net 
polarization (summed over the observable patch of the NS) may be significantly
reduced. However, this is incorrect \cite{heyl02,lai03b, vanadelsberg06}.
The reason is that, even in
vacuum\footnote{For photons with frequencies much higher than radio,
vacuum birefringence dominates over the plasma effect for all reasonable
magnetosphere plasma parameters \cite{wang07}.},
the photon polarization modes are decoupled near the NS surface
due to QED-induced birefringence, so that parallel-transport does not apply.

It is straightforward to obtain the observed polarized X-ray fluxes
(Stokes parameters) from the fluxes at the emission region, at least
approximately, without integrating the polarization evolution
equations in the magnetosphere \cite{lai03b,vanadelsberg06}.
For a given (small) emission region of projected
area (this is the area perpendicular to the ray at the
emission point --- General Relativistic light bending effect can be
easily included in this),  $\Delta A_\perp$, one need to know the
intensities of the two photon modes at emission, $I_\perp$ and
$I_\parallel$. In the case of thermal emission, these can be obtained
directly from atmosphere/surface models. As the radiation propagates
through the magnetosphere, the photon mode evolves adiabatically,
following the variation of the magnetic field, until the polarization
limiting radius $r_{\rm pl}$. This is where the two photon modes start
recoupling to each other, and is determined by the condition
$(\omega/c)\Delta n=2|d\phi_B/ds|$ (where $\Delta n$ is the 
difference of the indices of refraction of the two modes, $\phi_B$ specifies
the direction of $\vecB$ along the ray), giving
\begin{equation}
{r_{\rm pl}\over R_\ast}\simeq 70 \left( E_1 B_{\ast 13}^2P_1\right)^{1/6},
\end{equation}
where $R_\star$ is the NS radius and
$B_{\ast 13}$ is the polar magnetic field at the stellar surface
in units of $10^{13}$\,G [see ref.~\cite{vanadelsberg06}
for a more detailed expression).  Beyond $r_{\rm pl}$, the photon polarization
state is frozen. Thus the polarized radiation flux at $r\go r_{\rm pl}$ 
is $F_Q=(I_\parallel-I_\perp)\Delta A_\perp/D$, and 
$F_U\simeq F_V\simeq 0$\footnote{Note that $F_V$ is not exactly zero because 
of the neutron star rotation and because mode recoupling does not occur instantly
at $r_{\rm pl}$; see \cite{vanadelsberg06}.
}, where $D$ is the
distance of the source, $F_Q$ and $~F_U$ are defined in the coordinate
system such that the stellar magnetic field at $r_{\rm pl}$ lies in
the $XZ$ plane (with the $Z$-axis pointing towards the
observer). Since $r_{\rm pl}$ is much larger than the stellar radius,
the magnetic fields as ``seen'' by different photon rays are aligned and
are determined by the dipole component of the stellar field, one can
simply add up contributions from different surface emission areas to
$F_Q$ to obtain the observed polarization fluxes.

Note that the description of the polarization evolution in the last
paragraph is valid regardless of the possible complexity of the
magnetic field near the stellar surface. This opens up the possibility
of constraining the {\it surface} magnetic field of the neutron star
using X-ray polarimetry.  For example, the polarization light curve
(particularly the dependence on the rotation phase) depends only on
the dipole component of the magnetic field, while the intensity
lightcurve of the same source depends on the surface magnetic
field. On the other hand, the linear polarization spectrum (i.e., its
dependence on the photon energy) depends on the magnetic field at the
emission region \cite{lai03b,vanadelsberg06};
thus it is
possible that a NS with a weak dipole field ($\lo 7\times10^{13}$\,G)
may exhibit X-ray polarization spectrum characteristic of a
$B\go7\times10^{13}$\,G NS.

One complication arises from {\it Quasi-tangential} (QT) effect: As
the photon travels through the magnetosphere, it may cross the region
where its wave vector is aligned or nearly aligned with the magnetic
field (i.e., $\theta_B$ is zero or small). In such a QT region, the
two photon modes ($\parallel$ and $\perp$ modes) become (nearly)
identical, and can temporarily recouple, thereby affecting the
polarization alignment \cite{wang09}.
This QT effect gives rise to
partial mode conversion: after passing through the QT point, the
mode intensities change to ${\bar I}_\parallel\neq I_\parallel$ and
${\bar I}_\perp\neq I_\perp$. The observed polarization flux is then
${\bar F}_Q=({\bar I}_\parallel -{\bar I}_\perp)\Delta A/D\neq F_Q$.

In the most general situations, to account for the QT propagation
effect, it is necessary to integrate the polarization evolution
equations in order to obtain the observed radiation Stokes
parameters. However, Wang \& Lai (2009) showed that for generic
near-surface magnetic fields, the effective region where the QT effect
leads to significant polarization changes covers only a small area of
the neutron star surface. For a given emission model (and the size of
the emission region) and magnetic field structure, Wang \& Lai derived
the criterion to evaluate the importance of the QT effect. In the case
of surface emission from around the polar cap region of a dipole
magnetic field, they quantified the effect of QT propagation in detail
and provided a simple, easy-to-use prescription to account for the QT
effect in determining the observed polarization fluxes.  The net
effect of QT propagation is to reduce the degree of linear
polarization, so that ${\bar F}_Q/F_Q<1$, with the reduction factor
depending on the photon energy, magnetic field strength, geometric
angles, rotation phase and the emission area.  The largest reduction
is about a factor of two, and occurs for a particular emission size.
Obviously, for emission from a large area of the stellar surface, the
QT effect is negligible.

\section{Probing Axions with Polarized X-Rays}

The axion is a hypothesized pseudoscalar particle, introduced 
in 1980's to explain the absence of strong CP violation. 
The axion is also an ideal candidate for cold dark matter,
with the allowed axion mass
$m_a$ is in the range of $10^{-6}\lo m_a\lo 10^{-3}$~eV.

A general property of the axion is that it can couple to two photons
(real or virtual) via the interaction 
\begin{equation}
{\cal L}_{a\gamma\gamma}=-{1\over 4}\,g\,a\,F_{\mu\nu}{\tilde F}^{\mu\nu}=
g\,a\,{\bf E}\cdot\vecB, 
\end{equation}
where $a$ is the axion field, $F_{\mu\nu}$
(${\tilde F}^{\mu\nu}$) is the (dual) electromagnetic field strength
tensor, and $g$ is the photon-axion coupling constant. Accordingly, in
the presence of a magnetic field, a photon (the $\parallel$
component) may oscillate into an axion and vice versa.  
Exploiting such photon-axion
oscillation, various experiments and astrophysical considerations have been
used to put constraint on the allowed values of $g$ and $m_a$
\cite{raffelt08,cast09}.

Magnetic NSs can serve as a useful laboratory to probe axion-photon coupling.
Lai \& Heyl (2006) presented the general methods for calculating the axion-photon
conversion probability during propagation through a varying
magnetized vacuum as well as across an inhomogeneous
atmosphere. Partial axion-photon conversion may take place in the
vacuum region outside the NS. Strong axion-photon mixing occurs
due to a resonance in the atmosphere, and depending on the axion
coupling strength and other parameters, significant axion-photon
conversion can take place at the resonance. Such conversions may
produce observable effects on the radiation spectra and polarization
signals from the star. More study is needed in order to 
determine whether it is possible to separate out the photon-axion coupling 
effect from the intrinsic astrophysical uncertainties of the sources.
See ref.~\cite{lai06}
for more details and references.


\end{document}